\documentstyle[]{mn}

\begin{document}

\title[On the formation of black holes]
{Formation of massive stars and black holes in self-gravitating
AGN discs, and gravitational waves in LISA band.}

\author[Levin ]{Yuri Levin$^1$,$^2$ 
\\
$^{(1)}$601 Campbell Hall, University of California, Berkeley,
California, 94720\\
$^{(2)}$ Canadian Institute for Theoretical Astrophysics,
University of Toronto,  60 St.~George Street, Toronto, ON M5S 3H8,
Canada}
\date{printed \today}
\maketitle
\begin{abstract}
We propose a scenario in which massive stars
form at the outer edges of an AGN accretion disc.
We analyze the dynamics of a
disc forming around a supermassive black hole,
in which the angular momentum   is
transported by  turbulence induced
by the disc's self-gravity. 
We find that once the surface density of
the disc exceeds 
a critical value, 
the disc fragments into dense clumps. 
 We argue that the clumps accrete material from the remaining
disc and merge into larger 
clumps;
the upper mass of a merged clump is a few tens to
a few hundreds of solar mass. 
The biggest clumps collapse and form massive stars, which produce
few-tens-solar-mass black holes at the end of their evolution.


We construct a  model of
the AGN disc which includes extra heat sources from the
embedded black holes. 
If the embedded black holes can
accrete at the Bondi rate, then the feedback from accretion onto
the embedded black holes may stabilize
the AGN disc against the Toomre 
instability  for an interesting range of
the model parameters. 
By contrast, if the rate of accretion onto the embedded
holes is below the Eddington limit, then the extra heating
is insufficient to stabilize the disc.

The 
embedded black holes will 
interact gravitationally with the massive accretion disc
and be dragged
towards the central black
hole. Merger of a disc-born
black hole  with the central black hole will produce a burst
of gravitational waves. If the central black hole 
is accreting at a
rate comparable to the Eddington limit, the gas drag from the 
accretion disc will not alter significantly the dynamics
of the final year of merger, and the gravitational
waves should be observable  
by LISA. We argue that for a reasonable range of parameters
 such mergers will be detected
monthly, and that the gravitational-wave signal from these mergers is
distinct from that of other merger scenarios. Also, for some plausible 
black hole masses and accretion rates, the burst of gravitational waves
should  be accompanied by a detectable change in the  optical luminosity
of the central engine.

\end{abstract}

\section{introduction}

It is widely believed that self-gravitating accretion discs can form 
around supermassive black holes in AGNs.
Theoretical models show that the AGN  accretion discs must become
self-gravitating if they extend beyond a fraction of a parsec
away from a central black hole (Paczynski 1978, Kolykhalov and
Sunyaev 1980, Schlosman and Begelman 1987,
 Kumar 1999, Jure 1999, Goodman 2002). Self-gravitating discs are
unstable to fragmentation on a dynamical timescale; 
self-gravity of AGN accretion discs is a major 
issue in understanding how gas is delivered 
to the central black hole. It is likely that star formation
will occur in the outer parts of an AGN
  accretion disc (Kolykhalov and Sunyaev 1980, 
Schlosman and Begelman 1987).

Recently, some supporting evidence for this process was obtained from
observations of stellar velocities in 
our galactic center. Levin and Beloborodov (2003, LB) have analyzed the 3-D
velocity data of Genzel et.~al.~(2000), and have found that $\sim 80$
percent of massive stars from the  sample move in a thin
disc around the SgrA* black hole. LB's finding was confirmed by
Genzel et.~al.~(2003), who have used the most recent data from
Ott et.~al.~(2003). Genzel et.~al.~(2003) have also found evidence for
a second disc of stars which has a larger radius and thickness
than the disc in LB; the relative orientation of the two discs is nearly orthogonal. LB have argued that the stellar disc is a remant of a dense
self-gravitating accretion disc, much like what was already proposed
by Kolykhalov and Sunyaev in 1980 for the circum-black-hole accretion
discs in other galaxies (we quote from Kolykhalov and Sunyaev's
abstract: ``The fragmentation would produce around the black hole a ring of gas and
stars which would survive even after accretion onto the hole has ceased".) The
idea for star formation in the disc around SgrA* was originally proposed by
Morris et.~al.~(1999); in their
scenario the disc's self-gravity
does not 
play a central role. Instead, the compression of the disc material is achieved 
by radiation and wind pressure from the central engine. Morris et.~al.~(1999)
have identified the circumnuclear ring as the most
 likely source of supply for the disc
material. However, the circumnuclear ring is strongly misaligned with the disc
found by LB and by Genzel et.~al.~(2003). Thus, it seems more likely that the
gas in the disc would come from a disrupted molecular cloud on a plunging trajectory
(Sanders 1998).

A disc of stars could also be the remnant of a tidally disrupted young stellar
cluster (Gerhard 2001, Portegiese Zwart, McMillan, and Gerhard 2003,
McMillan and Portegiese Zwart 2003).
 This scenario becomes more credible if the dense cluster core
contains an intermediate-mass black hole (Hansen and Milosavlevic, 2003). The
gravitational pull from the latter 
would allow the core to stay intact until it comes within $0.1$pc
 from the SgrA*,
where the massive stars are observed to reside. 

Both the disc fragmentation and 
disrupted cluster scenarios have implications for the 
dynamics of the galactic nuclei and may provide channels for producing bursts
of gravitational waves detectable by LISA. In this paper we investigate the consequences
of disc fragmentation. 

The dynamics of the fragmenting disc is strongly affected by the feedback energy input
from the starburst.
Colliln and Zahn (1999) have conjectured that the feedback from this star formation 
will prevent the accretion disc from
becoming strongly self-gravitating. However,
Goodman (2003) has used general energy arguments to show
that the feedback from star formation is insufficient to
prevent an AGN disc  with the near-Eddington accretion rate
from becoming strongly self-gravitating
at a distance of $\sim$pc from the central black hole. This is
distinct from the case of galactic gas discs, for which  there
is evidence that the feedback from star formation protects  them  from
their self-gravitaty. 

In this paper we concentrate
on the physics of  the  self-gravitating disc and
make a semi-analytical estimate  of the possible
mass range of  stars formed in such  discs (Sections II
and III). 
 Our principal
conclusion is that the stars can be very massive,
up to hundreds of solar masses. These massive stars
evolve quickly and  produce stellar-mass black holes as the
end product of their evolution.

 Goodman (2003)
has mentioned the possibility that the AGN disc   is stabilized
by feedback from accretion onto black holes embedded
in the disc\footnote{Previously, 
Kolykhalov and Sunyaev (1980) have suggested
that the x-rays produced by accretion onto the embedded black holes
would be absorbed by the disc and reradiated in the infrared. The 
luminocity and spectra of the infrared radiation of the internally
heated AGN disc was recently computed by Sirko and Goodman (2003).}.
In Section IV, we investigate Goodman's conjecture  and
study whether the  feedback from accretion onto these black
holes may indeed stabilize a  disc around a supermassive black
hole, thus helping
to fuel bright AGNs. We find that if the embedded black holes
are accreting at a super-Eddington rate set by the Bondi-Hoyle
formula, then the extra heating they provide may be
able to stabilize the AGN disc. However, if their acretion rate
is no higher than the Eddington limit, than the AGN disc cannot
be stabilized against its own selfgravity, and one would need
some other source of heating to make the disc stable.
 
We then argue (Section V) that 
 the disc-born  black hole interacts gravitationally
with
the accretion disc,
and migrates inward 
on the timescale of $\sim 10^7$ years.  
The merger of the migrating black hole with the central
black hole will produce gravitational waves.
We show that for a broad range of
AGN accretion rates the final inspiral
is unaffected by gas drag, and therefore the
gravitational-wave signal should be detectable
by LISA.
 The  rate
of these mergers is uncertain, but if even a fraction of
a percent of the disc mass is converted into black holes
which later merge with the central black hole, then LISA should
detect monthly a signal from such a merger. The
final inspiral may occur close to the equatorial plane
of the central supermassive hole and is likely
to follow a quasi-circular orbit, which would make the
gravitational-wave signal distinct from those
in  other astrophysical merger scenarios.
If the disc-born black hole is sufficiently
massive, it will disrupt accretion flow in the disc
during the final year of its inpiral, thus making
an optical counterpart
to the gravitational-wave signal.

\section{physics of an accreting self-gravitating disc}
The importance of self-gravitating accretion discs
in astrophysics  has long been understood (Pazynsky 1978,
Lin and Pringle 1987). It was conjectured that the turbulence
generated by the gravitational (Toomre) instability may act as a 
source of viscosity in the disc. This viscosity  would both
drive accretion and keep the disc hot; the latter would act as
a  negative feedback for the Toomre instability
and would keep  the disc only marginally unstable.
Recently, there has been  big progress in our understanding
of the self-gravitating discs, due to a range of new and sophisticated
numerical simulations (Gammie 2001, Mayer et.~al.~2002,
Rice et.~al.~2003).
In our analysis, we shall rely extensively on these recent numerical
 results.

Consider an accretion disc 
which  is  supplied
by a gas infall. 
This  situation may arise when a merger or some other
major event in a  galaxy  delivers gas to the proximity of a supermassive
black hole residing in the galactic bulge. 
Let $\Sigma (r)$ be the surface density of the disc.
We follow the evolution of the disc
as $\Sigma(r)$ gradually increases due
to the  infall.

We begin by assuming that initially there is no 
viscosity mechanism, like Magneto-Rotational Instability
(MRI), to transport
the angular momentum and keep the disc hot\footnote{When
the disc begins to fragment, the viscousity due to self-gravity-driven
shocks exceeds the one due to MRI; see below. Therefore, while computing
the disc parameters at fragmentation, it is reasonable to ignore MRI}. 
This assumption
is valid when the ionization fraction of the gas in the
disc is low, i.e.~when the gas is 
far enough from a central source (about a thousand 
Schwarzchild radii from the supermassive black hole).
 We
also, for the time being, neglect irradiation from the central
source; this may be a good assumption if
a disrupted molecular cloud forms a disc
but the accretion onto the hole
has not yet begun.
 As will be discussed below,
irradiation is important for some regions
in the AGN discs we are considering. However,
as shown in the following subsection,
inclusion of irradiation or other
source of heating will only
strengthen the case for formation of massive stars.

We assume,
 therefore, that the forming disc  cools until
it becomes self-gravitating; this happens
when
\begin{equation}
Q={c_s\Omega\over \pi G \Sigma}\simeq 1.
\label{Q}
\end{equation}
Here $c_s$ is the isothermal speed of sound at the midplane
of the disc, and $\Omega$ is
the angular velocity of the disc.
Numerical simulations show that once the disc becomes self-gravitating,
turbulence and shocks develop; they
transport angular momentum and provide
heating which compensates the cooling of the disc
(Gammie 2001). We thus assume, in agreement with the simulations,
that when the disc exists, it is marginally
self-gravitating, i.e.~ $Q=1$. Then 
\begin{equation}
c_s={\pi G\Sigma\over \Omega},
\label{c_s}
\end{equation}
and
\begin{equation}
T\sim 2m_p c_s^2/k_B={2m_p\over k_B}\left(
   {\pi G\Sigma\over \Omega}\right)^2.
\label{T}
\end{equation}
Here  $T$ is the temperature in the
midplane of the disc, $m_p$ is the proton mass, and $k_B$
is the Boltzmann constant.

The one-sided flux from the disc
surface is given by the  modified Stephan-Boltzman
law:

\begin{equation}
F=\sigma T_{\rm eff}^4,
\label{flux1}
\end{equation}
where $T_{\rm eff}$ is the effective temperature. It is related
to the midplane temperature by
\begin{equation}
T_{\rm eff}^4\sim T^4  f(\tau)=f(\tau)\left({2m_p\over k_B}\right)^2
       \left(
   {\pi G\Sigma\over \Omega}\right)^8.,
\label{Trelation}
\end{equation}
where $\tau=\kappa \Sigma/2$ is the optical
depth of the disc; here $\kappa(T)$ is the opacity of the disc.
The function $f(\tau)=\tau$ for an optically thin disc, and
$f(\tau)=1/\tau$ for an optically thick disc.
We combine these two cases in our model by taking
\begin{equation} 
f(\tau)={\tau\over \tau^2+1}.
\label{ftau}
\end{equation}
We have used Eq.~(\ref{T}) in the last step of Eq.~(\ref{Trelation}).
The flux from the disc is powered by the accretion energy:
\begin{equation}
F={3\over 8\pi}\Omega^2 \dot{M}={9\over 8}\alpha\Omega c_s^2 \Sigma=
  {9\over 8}\alpha (\pi G)^2 {\Sigma^3\over \Omega}.
\label{flux2}
\end{equation}
Here $\alpha$ parametrizes viscous dissipation due to 
self-gravity (Gammie 2001); we have used Eq.~(\ref{c_s}) in the last step.
Using Eqs (\ref{flux1}), (\ref{flux2}), and (\ref{Trelation}), we can
express the viscosity parameter $\alpha$ as
\begin{equation}
\alpha={8\sigma \over 9}\left({2m_p\over k_B}\right)^4(\pi G)^6
       f(\tau){\Sigma^5\over \Omega^7}.
\label{alpha1}
\end{equation}
It is very important to emphasize that in this model
$\alpha$ is only a function of $\Sigma$ and $\Omega$:
the temperature in the midplane
is determined by Eq.~(\ref{T}), and this temperature
sets the opacity which in turn determines the
optical depth $\tau=\kappa \Sigma/2$. The  opacity
in the range of temperatures and densities
of interest to us is set by light scattering
off ice grains and, in some cases, by scattering off
metal dust. The relevant regimes are worked  out in the literature
on protoplanetary discs; we use the analytical fit
to the opacity (in cm$^2$/gm) from  the appendix of Bell and Lin, 1994.
\begin{eqnarray}
\kappa=0.0002\times T_K^2&\hskip 0.2in\hbox{for}&\hskip .2in T<166\hbox{K},\nonumber\\
\kappa=2\times 10^{-16}T_K^{-7}&\hskip 0.2in\hbox{for}&\hskip .2in166\hbox{K}<T<202\hbox{K}
\label{iceopacity}\\
\kappa=0.1\sqrt{T_K}&\hskip .2in\hbox{for}&\hskip .2in T>202\hbox{K}.\nonumber
\end{eqnarray}
In the first interval the opacity is due to icegrains; in the
second interval the icegrains evaporate, and the opacity
drops sharply with the temperature; in the third interval
(the highest $T$)  dust grains are the major source of the opacity.
Levin and Matzner (2003, in preparation) consider a more general temperature
regime and use the available opacity tables instead of relying
on the analytical fits.

From Eq.~(\ref{alpha1}) we see that
 as $\Sigma$ of the disc increases
due to the merger-driven infall, the effective viscosity will
reach $\alpha_{\rm crit}\sim 1$. At this stage the cooling time of the
disc becomes comparable to the orbital period. Gammie's   simulations
show that in this case the turbulence induced by self-gravity
is no longer able to keep the disc together, and the disc fragments.
Gammie's simulations give $\alpha_{\rm crit}\simeq 0.3$; the numerical
value we quote  disagrees with Gammie's,
but agrees with $\alpha_{\rm crit}$ quoted by  Goodman (2003) since like Goodman
we use isothermal speed of sound for $\alpha$-prescription.
Gammie's results, although obtained
for razor-thin discs, justify the key assumption of our model:
{\it the disc exists as a whole for $\alpha<\alpha_{\rm crit}$, and fragments
once $\alpha=\alpha_{\rm crit}$}.  Similar criterion
for the disc fragmentation was already used in Shlosman and
Begelman, 1987. 

We shall refer to the midplane
temperature and surface density of the marginally
fragmenting disc with $\alpha=\alpha_{\rm crit}$ as the critical
temperature $T_{\rm crit}$ and the critical surface density,
$\Sigma_{\rm crit}$.

We now find the critical surface density and midplane temperature
as a function of $\Omega$. We use Eq.~(\ref{T}) to express
$\Sigma$ as a function of $T$ and $\Omega$,  then substitute
this
function into Eqs.~(\ref{ftau}) and (\ref{alpha1}), and set
$\alpha=\alpha_{\rm crit}$. After simple algebra,
we obtain
\begin{equation}
\Omega^3+p\Omega=q,
\label{omega1}
\end{equation}
where
\begin{eqnarray}
p&=&2\left({\pi G\over \kappa(T_{\rm crit})}\right)^2{m_p\over k_B T_{\rm crit}},\nonumber\\
q&=&{32\sigma\over 9\alpha_{\rm crit}}{(\pi G)^2\over \kappa(T_{\rm crit})}
      (m_p/k_B)^2 T_{\rm crit}^2.\label{pq}
\end{eqnarray}

There is an analytical solution to Eq.~(\ref{omega1}):
\begin{equation}
\Omega=w-p/(3w),
\label{omega2}
\end{equation}
where
\begin{equation}
w=\{q/2+[(q/2)^2+(p/3)^3]^{1/2}\}^{1/3}.
\label{w}
\end{equation}

In Figure $1$ we make a plot of $T_{\rm crit}$
as a function of the orbital period, for concreteness
we set $\alpha_{\rm crit}=0.3$. The critically
self-gravitating disc is optically thin
if the second term of the LHS of Eq.~(\ref{omega1})
is dominant, and optically thick otherwise. This
can be expressed as a condition on the critical
temperature: the disc is optically thin if
\begin{equation}
T_{\rm crit}<14\hbox{K}\alpha_{\rm crit}^{2/15},
\label{tcrit1}
\end{equation}
and optically thick for higher critical temperatures.
The angular frequency above which the critically 
unfragmented disc becomes optically thick is 
\begin{equation}
\Omega_{\rm transition}\simeq 16.3\times 10^{-11}\hbox{sec}^{-1}.
\label{omegatrans}
\end{equation}

We use Eqs.~(\ref{c_s}) and (\ref{T}) to find the
critical surface density $\Sigma_{\rm crit}$, which is plotted
in Fig.~2,
and the scaleheight $h_{\rm crit}=c_s/\Omega$ 
of a marginally fragmenting disc. 
The Toomre mass $\bar{M}_{\rm cl}=\Sigma_{\rm crit}
h_{\rm crit}^2$ is the mass scale of the first clumps which
form in the first stage of fragmentation. In Fig.~3, we plot the Toomre
mass of the critically fragmenting disc as a function
of the orbital period.

The value of  $\bar{M}_{\rm cl}$ is not large enough
for the initial clump  to open
a gap in the accretion disc. The newly-born clump will
therefore accrete from the disc. The Bondi-Hoyle 
estimate of the accretion
 rate gives $\dot{M}_{\rm cl}\sim \Omega \bar{M}_{\rm cl}$,
i.e. we expect the mass of the new clump to grow on the dynamical timescale
until it becomes large  enough to open a gap in the gas disc. The upper limit
$\tilde{M}_{\rm cl}$
of this value is the mass which opens a gap in the original gas disc with 
$\Sigma=\Sigma_{\rm crit}$ just
before it fragments:
\begin{equation}
\tilde{M}_{\rm cl}\simeq \bar{M}_{\rm cl}
\left(40\pi\alpha_{\rm crit}\right)^{1/2}(r/h_{\rm crit})^{1/2};
\label{tildeM}
\end{equation}
see Eq.~(4) of Lin and Papaloizou (1986).
Once the gas is depleted from the disc, we expect the initial distribution
of the clump masses  to be concentrated between $\bar{M}_{\rm cl}$ and $\tilde{M}_{\rm cl}$.
The clump masses will evolve when clumps begin to merge with each other; this
 is addressed in  section III.

\paragraph{Effect of irradiation and other sources of heating}. So far 
in determining the structure of the self-gravitating disc, we have 
neglected external or internal
 heating of the disc. This is certainly a poor approximation
in many cases. Irradiation from AGN or surrounding stars, or feedback
from star formation inside the disc can be the dominant source of heating
of the outer parts of AGN discs (eg, Shlosman and Begelman 1987, 1989).
For example, iradiation from circumnuclear stars will keep the disc
temperature at a few tens of Kelvin, which is higher than the critical
temperature we obtained for a self-gravitating disc beyond $0.1$pc 
away from $10^7M_{\odot}$ black hole.  

However, extra heating will always work to increase the critical temperature at
which the disc fragmentation occurs.
Therefore, the values of the  critical surface density $\Sigma_{\rm cr}$,  scaleheight $h_{\rm cr}$, 
and the mass of the initial fragment $M_{\rm cl}$ obtained above should be treated
as lower bounds of what might be expected around real AGNs or protostars.
Higher values of these quantities would only strengthen main conclusions
of this paper. 
LB have found that when the rate of accretion $\dot{M}$ is constant throughout the disc, the
Toomre mass $\bar{M}_{\rm cl}$ is given by
\begin{eqnarray}
\bar{M}_{\rm cl}&=&1.8M_{\odot}\left({\alpha\over 0.3}\right)^{-1}\times\nonumber\\
                & &{\dot{M}c^2\over L_{\rm edd}}
\left({M\over 3\times 10^6M_{\odot}}\right)^{0.5}\left({r\over 0.2\hbox{pc}}\right)
^{1.5},
\end{eqnarray}
and that the gap-opening mass is 
\begin{eqnarray}
M_{\rm gap}&=&62M_{\odot}\left({\alpha\over 0.3}\right)^{-0.5}\times\\
& &{\dot{M}c^2\over L_{\rm edd}}
\left({M\over 3\times 10^6M_{\odot}}\right)^{0.5}\left({r\over 0.2\hbox{pc}}\right)^{1.5}
\left({r\over 10h}\right)^{0.5}.\nonumber
\end{eqnarray}
Here, $M$ is the black-hole mass, and  $r$ and $h$ are the radius and the scaleheight
of the disc.
Thus, even if the clumps do not merge with each other
and stop their growth at the gap-opening mass, the
mass of the formed stars will be biased towards the high-mass
end. In the next section we discuss the effects of clump mergers
and the mass range of stars born after the disc fragments.

\section{Evolution of the fragmented disc}

Gammie's simulations show that once the disc
fragments, the clumps merge and form significantly larger
objects. In fact, his razor-thin shearing box turned
into a single gas lump at the end of his simulation.

For merger to be possible,  the clumps should
not collapse into individual stars before they can
coalesce with each other. Let's check if this
is the case.

Consider a spherical nonrotating clump of radius $R$ and mass $M_{\rm cl}$.
First, assume that the clump is optically thin.
The  energy radiated  from the clump per unit time
is
\begin{equation}
W_{\rm cool}\sim \sigma T^4 R^2 \kappa(T)\Sigma\sim M_{\rm cl}\sigma T^4\kappa(T).
\label{Wcool}
\end{equation}
This radiated power cannot exceed the clumps gravitational binding energy
released in  free-fall time, $G^{1.5}M^{2.5}R^{-2.5}$. Together
with $\kappa(T)\propto T^2$ [since icegrains
dominate opacity for the optically-thin marginally
fragmenting disc--see Eq.~(\ref{tcrit1})],
this condition implies
that
\begin{equation}
T<\tilde{T}=T_0R^{-5/12},
\label{T1}
\end{equation}
where $T_0$ is a constant for the collapsing
optically thin clump. The temperature $\tilde{T}$ in
Eq.~(\ref{T1}) is less than the virial
temperature, which scales as $R^{-1}$.
Therefore, after the collapse commenses, the clump is not virialized while
it is optically thin. The temperature cannot
be much smaller than $\tilde{T}$, since otherwise
the cooling rate would be much smaller than the rate
of release of the gravitational binding energy,
and the gas would heat up by quasi-adiabatic compression.
The inequality in Eq.~(\ref{T1}) should be
substituted
by an approximate equality, and therefore we have during
optically-thin collapse
\begin{equation}
T\propto R^{-5/12}.
\label{T2}
\end{equation}
The optical depth scales as
\begin{equation}
\tau\propto R^{-11/3}.
\label{tau2}
\end{equation}
and hence rises sharply as the
clump's radius decreases; as the clump shrinks
it  becomes optically
thick\footnote{The contraction of an optically thin clump
may be complicated by subfragmentation, since the
Jean's mass for such clump scales as $R^{7/8}\propto\tau^{-0.23}$.
We suspect that in most cases the clump becomes optically
thick before it subfragments, since the Jean's mass has a slow
dependence on the optical depth. However, only detailed numerical simulations
can resolve these issues.}. It is possible to show that once the clump
is optically thick, it virializes quickly with it's temperature
$T\propto R^{-1}$. For $\kappa\propto T^2$ (icegrains),
the cooling time of an optically thick clump scales with the clump radius as
\begin{equation}
t_{\rm cool}\propto R^{-3}.
\label{tcool}
\end{equation}
The characteristic timescale for the clump
to collide with another clump scales with the 
clump radius as 
\begin{equation}
t_{\rm collision}\propto R^{-2}.
\label{tcoll}
\end{equation}
From Eqs.~(\ref{tcool}) and (\ref{tcoll}),
we see that the collision rate decreases less
steeply
than the coolling rate as a function of the radius
of an optically thick clump. Therefore, merger
can be an efficient way of increasing the clump's mass.

This conclusion is no longer valid when the temperature
of the clump becomes  larger than $\sim 200$K; then
the opacity is dominated by metal dust with $\kappa\propto T^{1/2}$.
In this case the cooling time scales as $t_{\rm cool}\propto R^{-1.5}$.
The collision timescale increases faster than the cooling time
as the clump shrinks, and naively one would expect
that mergers may not be efficient in growing
the clump masses.

However,
 we have neglected  the rotational
support within a clump. Each clump is
initially rotating with angular frequency 
comparable to the clump's inverse dynamical timescale;
for example, in a Keplerian disc each clump's  
initial angular velocity is  $\sim \Omega/2$.
Therefore each clump will shrink and collapse
into a rotationally supported  disc, and the size
of this disc is comparable to the size of the original
clump (this picture seems to be in agreement with
Gammie's simulations).
Thus rotational support generally 
slows down the collapse of an individual fragment 
and makes mergers between diferent fragments
to be efficient.

 Magnetic braking is one of
the  ways for the clump to  lose its  rotational support\footnote{Another way
is via collisions with other clumps.}
(see, e.g., Spitzer 1978). 
One generally expects a horizontal magnetic field to be present
in a differentially rotating disc due
to the MRI (Balbus and Hawley, 1991). Ionization fraction
in the disc
is expected to be small,
so the magnetic field is saturated at a subequipartition
value $B=\beta B_{\rm eq}$, with $\beta<<1$. Horizontal magnetic field will couple 
inner and outer parts of the differentially rotating clump
on the Alfven crossing timescale $t_{\rm alfven}\sim t_{\rm dynamical}/\beta$,
and the collapse will proceed on this timescale as well.

What is the maximum mass that the clump can achieve?
This issue has been analyzed for the similar situation
of a protoplanetary core accreting from a disc of
planetesimals (Rafikov 2001 and references therein).
The growing clump cannot accrete more mass than is 
present in it's ``feeding annulus''. This gives the maximum 
``isolation''  mass of a clump:
\begin{equation}
M_{\rm is}\sim {(2\pi r^2 \Sigma_{\rm crit})^{3/2}\over
         M^{1/2}}=2\pi\sqrt{2}\bar{M}_{\rm cl}(r/h_{\rm crit})^{3/2},
\label{Mis}
\end{equation}
where, as above, $\bar{M}_{\rm cl}=\Sigma_{\rm crit}h_{\rm crit}^2$
is the mass scale of the first clumps to form
from a disc; see, e.g.,  Eq.~(2) of Rafikov (2001). However, numerical
work of Ida and Makino (1993) and analytical calculations of
Rafikov (2001) indicate that the isolation mass may be hard
to reach. The consider a massive body moving on a circular orbit
through a disc of gravitationally interacting particles, and
they find that when the mass of the body exceeds some critical
value, an annular gap is opened in the particle disc around
the body's orbit.  We can idealize a disc consisting of fragments
as a disc of particles of a typical fragment mass $M_{\rm fr}$.
Once a growing clump opens a gap in a disc of gravitationally
interacting fragments, the clump's growth may become quenched.
 This gap-opening  mass of the
clump $M_{\rm gap}$  is given by Eq.~(25) of
Rafikov (2001):
\begin{equation}
{M_{\rm gap}\over M_{\rm is}}={I\over 2^{7/6}\pi^{1/2}}
\left({M_{\rm fr}\over \Sigma r^2}\right)^{1/3} \left(
{M\over \Sigma r^2}\right)^{1/2}.
\label{rafikov25}
\end{equation}

 We use the numerical factor
$2^{-7/6}\pi^{-1/2}I=1.5$ appropriate for thin discs. 
By taking $Q=1$ we get 
\begin{equation}
M_{\rm gap}\simeq 14 M_{\rm fr} (M_{\rm fr}/\bar{M}_{\rm cl})^{1/3}(r/h_{\rm crit}).
\label{mgap}
\end{equation}
In Figure 4  the masses $M_{\rm is}$
and $M_{\rm gap}$ are plotted
as a function of radius for a $3\times 10^6M_{\odot}$
black hole; when we calculate  $M_{\rm gap}$ we
conservatively set $M_{\rm fr}=\bar{M}_{\rm cl}$ and not
to the larger value $\tilde{M}_{\rm cl}$. It is likely that the most massive 
clumps will reach $M_{\rm gap}$,
but it will  be more difficult to form a clump
with the mass $M_{\rm is}$.

From Fig.~$4$ we see that the most 
massive clumps can reach tens hundereds of solar masses.
The maximum mass would be even larger if we included
the heating of the disc by external irradiation or 
internal starburst.
It is plausible that these  
 very massive clumps will form   massive stars;
the masses of the stars may be comparable to the
masses of the original clumps; see McKee and Tan (2002)
and references therein. Stars with masses of a few tens
of solar masses 
will produce black holes as the end product
of their rapid ($<10^6$yr) evolution; the characteristic
mass of these black holes is believed 
to be around $10M_{\odot}$. A recent work by Mirabel and
Rodrigues (2003)
shows that the  black holes whose projenitors have
masses $>40M_{\odot}$ do not recieve a velocity kick at their birth.
Thus, the disc-born black holes are likely to remain embedded
in the disc.

\section{AGN disc with embedded stellar-mass
         black holes}
Once formed, the stellar-mass black holes will
dramatically affect the dynamics of the AGN disc.
We assume that the disc is replenished due to continuous
infall, and in this section we analyze a steady-state
AGN disc which is heated by the energy released
due to the the accretion onto stellar-mass black
holes embedded in the disc. 

We make an anzatz that the heated disc is optically
thick and is supported by the radiation pressure.
Once we obtain the solution for the structure
of the disc, we will derive the regime of
validity for our anzats.
We thus have for the sound speed in the disc midplane: 
\begin{equation}
c_s^2 ={aT^4\over 3\rho}\sim {2aT^4\over 3\Sigma}{c_s\over \Omega},
\label{csmid}
\end{equation}
and hence
\begin{equation}
c_s\sim{4\kappa F\over 3\Omega c}.
\label{csmid1}
\end{equation}
Here we have used the expression for the flux
through the disc face, $F\sim caT^4/(2\Sigma \kappa)$.

For simplicity, we assume that there are $N_{\rm bh}$ black
holes of mass $M_{\rm bh}$ embedded in a disc within
radius $r$ from the central hole of mass $M$.
For this ``planetary'' system to be stable,
$N_{\rm bh}$ cannot be much bigger than $(M/M_{\rm bh})^{1/3}$,
i.e. the embedded holes cannot be within each other's
Hill spheres. 

To make further progress, we need to know the rate of accretion
onto the black holes embedded in the disc. Because
the density of the disc material is high, the well-known Bondi-Hoyle
formula gives a super-Eddington accretion rate; it is not clear
at this moment how feasible this is (but see e.g.~Begelman 2002).
We therefore consider below two separate models: the ``Bondi-Hoyle''
model, in which the embedded black holes accrete at the Bondi-Hoyle
rate, and the ``Eddington'' model, in which the accretion onto
the embedded holes occurs at the Eddington limit.

\subsection{The ``Bondi-Hoyle'' model}.
In this model, the accretion rate for the embedded black home of
mass $M_{\rm bh}$ is given by the Bondi-Hoyle formula, 
\begin{equation}
\dot{M}_{\rm bh}\simeq 4\pi r_{\rm bondi}^2\rho c_s\sim
2\pi r_{\rm bondi}^2 \Sigma \Omega.
\label{dotMbh}
\end{equation}
For the latter expression to be true, the 
Bondi radius $r_{\rm bondi}=GM_{\rm bh}/c_s^2$ must
be less than the scaleheight of the disc; this condition will 
 be checked once the structure of the disc is worked out. 

We assume that the fraction $\epsilon$ of 
the rest-mass energy of the accreted material
goes into heating of the disc\footnote{We envisage that each hole
powers a minijet which carries a significant
fraction of the accretion energy.}, is thermalized, and is eventually
radiated as flux $F$ from the disc surface:
\begin{equation}
F=\epsilon {N_{\rm bh}\over 2\pi r^2} \dot{M}_{\rm bh}c^2.
\label{fluxbh}
\end{equation}
Finally, the equation for
the central hole's accretion rate
\begin{equation}
\dot{M}=3\pi\alpha\Sigma c_s^2/\Omega,
\label{dotMcentral}
\end{equation}
together with the Equations (\ref{csmid1}), (\ref{dotMbh}), and (\ref{fluxbh})
 allow us to find the structure of the disc once the quantities
$M_{\rm bh}$, $N_{\rm bh}$, $M$, $\dot{M}=\dot{m}\dot{M}_{\rm edd}$, 
$r$, $\alpha$, $\kappa$, and $\epsilon$ are fixed. After some
algebra, we find the expression for the Toomre parameter
of the disc
\begin{equation}
Q\sim 10 {\alpha_{0.3}^{4/7}(\kappa/\kappa_0)
({\epsilon/\dot{m}})^{4/7}
\left({
M_{\rm bh}/10M_{\odot}}\right)^{5/7}\over M_7^{3/14} 
({r/0.1\hbox{pc}})^{3/2}}.
\label{Qbh}
\end{equation}
We see that a typical AGN disc can be stabilized
by feedback from embedded  black holes out
to a radius of about a parsec. Here $\kappa_0$ is the 
opacity for Thompson scattering, and $N_{\rm bh}$ was set
to its upper limit, $(M/M_{\rm bh})^{1/3}$. 

Recently, Sirko and Goodman (2002) analyzed SEDs from
AGN discs in which $Q=1$ is enforced by
the unspecified heating sources. They have shown that current observations
limit such self-gravitating discs to be no larger than $\sim 10^5$ 
Schwartzchild radii, about $0.1$pc for $10^7M_{\rm odot}$ black hole.
This motivates our  normalization for the radius used in Eq.~(\ref{Qbh}).

We now
check the assumptions which were used in calculating
the disc structure:
\begin{equation}
{p_{\rm rad}\over p_{\rm gas}}\sim 1.5*10^2{(\kappa/\kappa_0)^{1/4}
M_7^{5/56}(M_{\rm bh}/10M_{\odot})^{15/28}\over\alpha_{0.3}^{1/14}
(\epsilon/\dot{m})^{1/14}(r/0.1\hbox{pc})^{3/8}}
\gg 1,
\end{equation}
so the disc is radiation-pressure dominated;
\begin{equation}
\tau\sim 10^2 {(\dot{m}/\epsilon)^{5/7}
M_7^{41/42}\over (M_{\rm bh}/10M_{\rm odot})^{10/21}
(r/0.1\hbox{pc})^{1/2}\alpha_{0.3}^{5/7}}\gg 1,
 \end{equation}
hence the disc is optically thick;
and 
\begin{equation}
r_{\rm bondi}/h=2*10^{-3}{\alpha_{0.3}^{3/7}\epsilon^{3/7}(M_{\rm bh}/10M_{\odot})^{2/7}
\over M_7^{2/7}\dot{m}^{3/7}}\ll 1,
\end{equation}
which gives some credibility to
the Bondi-Hoyle estimate of the
accretion rate onto the disc-born
black hole. 

The expressions for other useful
quantities characterizing the disc
are  given below:
\begin{eqnarray}
h/r&\sim&0.08{(M_{\rm bh}/10M_{\odot})^{5/21}(\dot{m}/\epsilon)^{1/7}
\over \alpha_{0.3}^{1/7}M_7^{5/21}}\label{hr}\\
c_s&\sim&10^7\hbox{cm/s}
{(M_{\rm bh}/10M_{\odot})^{5/21}M_7^{11/42}
\over (\epsilon/\dot{m})^{1/7}\alpha_{0.3}^{1/7}(r/0.1\hbox{pc})^{1/2}}  ,\label{csvalue}\\
{N_{rm bh}\dot{M}_{\rm bh}\over\dot{M}}&\sim&0.08{(M_{\rm bh}/M_{\odot})^{5/21}\over
\alpha_{0.3}^{1/7}\epsilon^{1/7}\dot{m}^{6/7}M_7^{5/21}}.\label{ratiovalue}
\end{eqnarray}
The last quantity is the characteristic fraction of the
disc mass which gets converted into   embedded black holes.

\subsection{The ``Eddington'' model}
The Eddington acretion rate for the embedded black hole
is 
\begin{equation}
\dot{M}_{\rm bh}={4\pi GM_{\rm bh}\over\epsilon_a\kappa_0 c},
\label{meddington}
\end{equation}
where $\epsilon_a$ is the radiative efficiency of an
accretion flow.
We analyze the structure of the disc by repeating the
steps outlined in the previous section, and by 
adopting the upper limit $(M/M_{\rm bh})^{1/3}$ for the
number of the black holes embedded in the disc. We find
\begin{equation}
Q\sim 3*10^5\alpha_{0.3}(M_{\rm bh}/M)^2 (\kappa/\kappa_0)^3,
\label{Qeddington}
\end{equation}
and 
\begin{equation}
h/r\sim 2(\kappa/\kappa_0)(M_{\rm bh}/M)^{2/3}.
\label{hreddington}
\end{equation}
However an embedded black hole
 will open a gap if its mass is 
bigger than 
\begin{equation}
M_{\rm bhgap}\simeq\sqrt{40\alpha}(h/r)^{2.5}M,
\label{mbhgap}
\end{equation}
see Eqs.~(4) of Lin and Papaloizou (1986). 
Using Eq.~(\ref{hreddington}), we see that
\begin{equation}
M_{\rm bhgap}/M_{\rm bh}\sim 20(M_{\rm bh}/M)^{2/3}\ll 1.
\label{mbhgapmbh}
\end{equation}
The inequality above holds unless we allow
the individual masses of the embedded holes be
of order of tens of thousands solar masses.
Therefore it is impossible to maintain a stable disc by using 
the feedback from Eddington-limited black holes: either these
holes are not massive enough to provide sufficient feedback,
or they are so massive that they open gaps in and become decoupled
from the disc. An unsupported disc would fragment completely and
the fuel supply through the disc to the central engine
would be stopped.

\section{Merger of the central black hole
and the disc-born black hole.}
It is likely that the newly-born black hole
in the disc  inspirals towards the central
black hole. We imagine that 
 the stellar-mass black hole
is embedded into a massive accretion disc which
forms due to continuing infall of gas from the
galactic bulge, after the black hole is born. If the
black hole opens a gap in the disc, it will move
towards the central black hole together with the disc
(type-II migration; Gould and Rix 2000, and Armitage
and Natarajan 2001). The timescale for such inspiral
is the accretion timescale,
\begin{equation}
t_{\rm inspiral}\sim 
10^6\hbox{yr}{M_7^{-1/2}\over \alpha_{0.1}}\left(
{0.1\hbox{pc}\over r}\right)^{-3/2}\left({r\over 30h}\right)^2.
\label{typeII}
\end{equation}
If on the other hand, the black hole is not massive 
enough to open the gap, it will migrate inwards by
exciting  density waves in the disc (type-I migration).
The speed of this inward drift is given by
\begin{equation}
v_{\rm in}\sim 2\beta {G^2M_{\rm bh}\Sigma h\over c_s^3r};
\label{typeIv}
\end{equation}
cf.~Eqs.~(9), (B4) and (B5) of Rafikov (2002), and Ward (1986).
Here $\beta$ is a numerical factor of order $5$ for  $Q\gg 1$, and
can be significantly larger for $Q\sim 1$. For a self-gravitating
disc with $Q\sim 1$, we find    
\begin{equation}
t_{\rm inspiral}\sim 10^7\hbox{yr}{5\over\beta}{30h\over r}{10M_{\odot}
\over M_{\rm bh}}\left({r\over 0.1\hbox{pc}}\right)^{1.5}M_7^{0.5}.
\label{typeI}
\end{equation}

What determines whether the gap is open or not is whether a 
stellar-mass black hole has time to accrete a few thousand solar
masses of gas, which would put it above the gap-opening threshold
for the typical disc parameters. The Eddington-limited accretion
occurs on a timescale of $\sim 10^8$yr, much longer than the
characteristic type-I inspiral time. Thus, in this case the black
holes don't accrete much on their way in.
On the other hand, the Bondi-Hoyle formula predicts the mass e-folding
timescale of a few hundred years. Thus, if the black hole
is allowed to accrete at the Bondi-Hoyle rate, its mass
increases rapidly until it opens a gap in the disc. Then, the
inspiral proceeds via type-II migration.

From Eqs.~(\ref{typeII}) and (\ref{typeI}) we see that the embedded
black hole experiencing type-I or type-II migration
would merge with the central black hole on the timescale $10^6$---$10^7$
years, shorter than the typical timescale of an AGN activity.
Thus it is plausible that the daughter  disc-born
black hole is 
brought towards the parent central black hole; the mass of the
daughter black hole  might grow significantly on the way in. Gravitational radiation
will eventually become the dominant mechanism driving the inspiral,
and the final merger will produce  copious amount of gravitational
waves. In the next subsection we show that
these waves  are detectable by LISA for a broad range of
the black hole masses and the disc accretion rate.

\subsection{Influence of the  accretion disc on the inspiral signal 
as seen by LISA}

It is realistic to expect that LISA would follow the last year of the inspiral
of the disc-born black hole into the central black hole.
Generally, one must develop a set of templates which
densely span the parameter space of possible inspiral signals. In order
for the final inspiral to be detectable, one of the
templates must follow the signal with the phaseshift between the two not exceeding a fraction
of a cycle. Therefore, if the drag from the disc
will alter the waveform by a fraction of a cycle over the signal integration
time (e.g., $1$ year),   detection of the signal  with high
signal-to-noise ratio will become problematic.
Below we address the  influence of the accretion disc  on
the final inpiral waveform.

The issue of gas-drag influence on the LISA signal was first
addressed by Narayan (2000); see also 
Chakrabarti (1996). Narayan's analysis is directly applicable
to low-luminocity non-radiative quasi-spherical accretion flows,
which might exist around supermassive black holes when the accretion rate
is $<0.01$ of the Eddington limit. Narayan concluded that non-radiative
flows will not have any observable influence on the gravitational-wave signals
 seen by LISA. Below we extend Narayan's analysis to the case of
radiative disc-like flows with high accretion rate, which are likely
to be present in high-luminocity AGNs.

Consider a non-rotating central black hole of mass $M_6\times 10^6M_{\odot}$,
accreting at a significant fraction $\dot{m}$ of the
Eddington rate, $\dot{M}=\dot{m}\dot{M}_{\rm edd}$.
The accretion disc in the region of interest ($<10 R_S$, where
$R_s$ is the Schwartzchild radius of the central black hole) is radiation-pressure
dominated, and the opacity $\kappa=0.4\hbox{cm}^2/\hbox{g}$ is due
to the Thompson scattering. By following the standard thin-disc 
theory\footnote{The disc is no longer thin close
to the central black hole, however our estimates of
the disc structure should be correct to an order of magnitude.}
(Shakura and Sunyaev, 1973), we get
\begin{equation}
\Sigma(r)={64\pi c^2\over 27\alpha\Omega\kappa^2\dot{M}}\sim 4\hbox{g}/\hbox{cm}^2{\epsilon\over \alpha \dot{m}}\left(
        {r\over r_s}\right)^{3/2},
\label{sigmabh}
\end{equation}
where $\epsilon$ is the efficiency with which the accreted mass converts into
radiation, and $r_s$ is the Schwartzchild radius of the central black hole;
and 
\begin{equation}
c_s={3\dot{M}\Omega\kappa\over 8\pi c}.
\label{csbh}
\end{equation}

The disc black hole in orbit around the
central black hole  excites density
waves in the disc (Goldreich and Tremaine, 1980); these waves carry angular momentum flux
\begin{equation}
F_0\sim \left(GM_{\rm bh}\right)^2 {\Sigma r \Omega\over c_s^3};
\label{fluxwaves}
\end{equation}
here, as before,  $M_{\rm bh}$ is the mass of the orbiting disc black hole.
Ward (1987) has argued that the torque acting on the
orbiting body is $dL_{\rm dw}/dt \sim (h/r)F_0$. We can compute the
characteristic timescale for the orbit evolution due to the 
density-waves torque:
\begin{equation}
t_{\rm dw} = {L\over dL_{\rm dw}/dt}={1\over \Omega}{M\over M_{\rm bh}}{M\over \Sigma r^2}\left({h\over r}\right)^2.
\label{torquewaves}
\end{equation}
One must compare $t_{\rm dw}$
to the timescale $t_{\rm gw}$ of orbital evolution due to gravitational-radiation-reaction tourque:
\begin{equation}
t_{\rm gw}=8t_{\rm m}={5\over 8}{cr_s^2\over GM_{\rm bh}}\left({r\over r_s}\right)^4.
\label{tgw}
\end{equation}
Here, $t_m$ is the time left before the disc and central holes merge, i.e.~the
integration time for the LISA signal.
Optimistically we could expect to follow the LISA signal to $0.1$ of a cycle
(Thorne 1999). Therefore, if $q=10 n_{\rm m} t_{\rm gw}/t_{\rm dw}$ is less than
unity, the disc drag does not impact detection of the final inspiral; cf.~Eq.~(16)
of Narayan (2000). Here $n_{\rm m}=\Omega t_{\rm gw}/(5\pi)$ is the number
of cycles the disc black hole will make before merging with the central black hole.
Using Eqs.~(\ref{sigmabh}), (\ref{csbh}), (\ref{torquewaves}), and (\ref{tgw}), we 
get
\begin{equation}
q\simeq 2\times 10^{-7}{\epsilon_{0.1}^3\over \dot{m}^3 \alpha_{0.1}} 
{(M_{\rm bh}/10M_{\odot})^{13/8}\over M_6^{13/4}}
t_{m, yr}^{21/8},
\label{q}
\end{equation}
where $\epsilon=0.1\epsilon_{0.1}$, $\alpha=0.1\alpha_{0.1}$, and
$t_m=1\hbox{yr}*t_{m, yr}$.
We see that for a large range of parameters $q<1$, and the disc drag does not
influence the inpiral signal. However, note that $q$ is a very sensitive
function of $\dot{m}$ and $M_{\rm bh}$. For instance, the disc will influence significantly
an inspiralling $100M_{\odot}$ black hole when the accretion rate is down to a few
percent of the eddington limit. One then needs to reduce the influence
of the disc on the LISA signal
 by choosing to observe the smaller portion of the final
inspiral, i.e.~by choosing a smaller integration time $t_m$.

There is another important source of drag experienced by the inpiralling
black hole;  it was first analyzed by Chakrabarti (1993, 1996). Generally,
there is a radial pressure gradient in an accretion disc; this pressure 
gradient makes the asimuthal velosity of the disc gas  slightly different
from a   velocity of the test particle on a circular orbit at the same radius.
Therefore the  inspiralling black hole will experience a 
head wind from a gas in the accretion disc;  by accreting gas
from the disc the black hole will experience  the braking torque which
will make it lose it's specific angular momentum. This tourque is given 
by 
\begin{equation}
\tau_{\rm wind}=\dot{M}_{\rm bh}\Delta v r,
\label{tauwind}
\end{equation}
where $\dot{M}_{\rm bh}$ is rate of accretion onto the inspiraling 
black hole from the disc, and $\Delta v\sim c_s^2/v_o$ is the speed of the 
headwind experienced by the black hole moving with the orbital speed $v_o$.
We assume that the inspiraling hole accretes with the Bondi-Hoyle rate,
\begin{equation}
\dot{M}_{\rm bh}\simeq \pi\rho (GM_{\rm bh})^2/c_s^3,
\label{mdisc}
\end{equation}
where $\rho=\Sigma/h=\Sigma\Omega/c_s$ is the density of the ambient
disc gas.  By using the last equation in Eq.~(\ref{tauwind}), we obtain
the expression for the tourque $\tau_{\rm wind}$, which turns out
to be the same as the tourque from the density waves, up to the numerical
factor between $1$ and $10$. We see therefore that inclusion of Chakrabarti's
``accretion'' torque is important  for the detailed analysis,
but does not qualitatively change our conclusions.    

Can the orbiting  black hole open  
a gap in the disc during it's final inspiral?
In order to overcome the viscous stresses which oppose  opening
the gap, the mass of the orbiting black hole should be greater than
the  threshold value
given by Eq.~(\ref{mbhgap}).
The radius $r_{\rm in}$ from which the inspiral begins is 
given by 
\begin{equation}
r_{\rm in}=4(M_{\rm bh}/10M_{\odot})^{1/4}M_6^{-1/2}t_{m, yr}^{1/4}r_s,
\label{rin}
\end{equation}
and the scaleheight of the disc is radius-independent in the radiation-dominated
inner region (this is true only if you treat the central black hole as a Newtonian
object):
\begin{equation}
h={3\dot{m}\over 4\epsilon}r_s.
\label{hin}
\end{equation}
The gap will be open in the disc, therefore, if the mass of
the inspiraling black hole exceeds
\begin{equation}
m_{\rm gap}\sim \dot{m}^{20/13}M_6^{-2/13} 10^4M_{\odot}.
\label{mvgap1}
\end{equation}
During the final year of merger of the   $10M_{\odot}$ disc-born black hole
and  the $10^6M_{\rm odot}$ central black hole, the gap will
be open if the accretion rate onto the central
black hole is about five  percent of the Eddington limit.
Such gap-opening merger can result in sudden changes in the
AGN luminosity and produce an optical counterpart to the 
gravitational-wave signal. 

\subsection{The merger event rate as seen by LISA}
The details of black-hole formation and evolution in the disc are 
uncertain, and   it seems impossible to make a reliable estimate
of an event rate for LISA from this channel of black-hole mergers.
Nonetheless, from Eq~(\ref{ratiovalue}) we see
that in order for  an extended accretion disc
in a bright AGN to be stable,
a significant mass fraction of the material
acreted by the cental black hole
must go  into the stellar-mass black holes embedded
in the disc. 
It is worth working through a simple example to
show that the merger of disc-born holes with the central hole
may  be an important source for LISA.

Studies of integrated light coming from Galactic Nuclei show
that the supermassive black holes acquire significant part and
perhaps almost all of their mass via accretion of gas;
see, e.g., Yu and Tremaine (2002) and references therein. Assume,
for the sake of our example, that a mass fraction $\eta$ of this gas
is converted into $100M_{\odot}$ black holes on the way in,
and that all of these black holes eventually merge  with the central
black hole. LISA can detect such mergers to $z=1$ if the central
black hole is between $10^5$ and $10^7$ solar masses (one needs to assume
that the central black hole is rapidly rotating at the high-mass end
of this range). The mass density of such black holes in the
local universe is $\sim 10^5 M_{\odot}/\hbox{Mpc}^3$ (Salucci et.~al.~1999).
We can estimate, using Figure $3$ of Pei et.~al.~(1995),
that about $10$ percent of  integrated radiation from Galactic Nuclei
comes from redshifts accessible to LISA, $z<1$. This implies that
 supermassive black holes acquired $10$ percent of their mass
at $z<1$. We shall therefore assume that $10$ percent of the mass of the
black holes in LISA mass range was accreted at $z<1$. This implies that
there were $\sim 100\eta=\eta_{0.01}$ mergers per mpc$^3$ which are potentially detectable
by LISA. When we multiply this by volume out to $z=1$ and divide it by
the Hubble time, we get an estimate of the LISA event rate from 
such mergers,
\begin{equation}
dN/dt_{\rm toymodel}\sim 10\eta_{0.01}/\hbox{yr}.
\label{rate}
\end{equation}
The estimate above should be treated as an illustration of importance
for our channel of the mergers, rather than as a concrete prediction for LISA.

Currently, it is not
known how to compute well the gravitational waveform produced
by an inpiral with an arbitrary eccentricity and inclination relative to the central 
black hole. This might pose a great challenge to the LISA data analysis.
Indeed, in the currently popular  astrophysical scenario, 
the stellar-mass black hole gets captured on a highly eccentric
orbit with arbitrary inclination relative to the
 central black hole (Siggudrson and Rees, 1997).
By contrast, in our scenario the inpiral occurs in the equatorial plane, and the signal is well
understood (Hughes 2001 and references therein). Hence the
template for detection is readily available, and
the merger channel we consider has a clear observational
signature which distinguishes
it from other channels. It is an open question whether the inpiraling
hole can acquire high eccentricity by ineracting with the disc or with other orbiting masses
(Goldreich and Sari 2003, Chiang et.~al.~2002); it seems likely that at least in some
cases interaction with the disc will act to circularize the orbit. Finn and Thorne (2000) have
performed a detailed census of parameter space for circular-orbit  inspirals as seen by LISA.  

\section{conclusions}
In this paper,
we rely on  numerical simulations by Gammie (2001)
to develop a  formalism for self-gravitating 
thin discs which are gaining mass by continuous infall.
We present a  way to calculate the critical
temperature, surface density, and scaleheigh of the
disc just prior to fragmentation. Our formalism naturally 
includes both 
optically thin and optically thick discs.

We then speculate on the outcome of the nonlinear physical
processes which follow fragmentation: accretion and merger
of smaller fragments into bigger ones. We find an upper
bound on the mass of final, merged fragments, and we give
a plausibility argument that some fragments will
indeed reach this upper bound. We thus predict that
very massive stars of tens or even hundreds of solar masses
will be produced in self-gravitating discs around supermassive
black holes. The end product of fast evolution of these massive
stars will be stellar-mass black holes. 

Finally, we make an  argument that 
the disc-born black holes in AGNs  find a way to merge
with  central black holes. We consider a purely toy-model
example of what the rate of such mergers might be, as seen
by LISA; we illustrate this rate might  be high enough 
to be  interesting 
for future gravitational-wave (GW) observations.
The  GW signal from this merger channel is
distinct from that of other channels, and can be readily
modeled using our current theoretical understanding of the 
final stages of the inspiral driven by gravitational-radiation reaction.
We 
 show that for a broad range
of accretion rates and black-hole masses
the drag from accretion disc will not be large enough
to pollute the signal and make inspiral template invalid.
In some cases, the inspiralling
hole will open a gap in the accretion disc close to
the central black hole, thus producing  a possible optical
counterpart for the gravitational-wave burst generated    
by  the merger. 

\section{acknowledgements}
I have greatly benefited from discussions
with Chris Matzner throughout this project.
He has 
suggested the model for the heated disc
in section IV, and has independently verified
the formulae in that section. Prof.~Matzner
has
graciously declined to be a co-author of this paper.
I have learned a lot about
accretion discs from  Eliot Quataert.
After the bulk of this project was completed, I
have found out that Jeremy Goodman and Jonathan Tan have pursued
similar line of research; our open exchange of ideas
is much appreciated.
I also thank  Andrei Beloborodov,  
 Andrew Melatos, Norm Murray, Frank Shu, Anatoly Spitkovsky, Kip Thorne,
 and Andrew Youdin for  discussions
and insights.  Sarah Levin has contributed to the
clarity of the prose. This research  was supported by TAC at UC Berkeley,  by
NSERC at CITA, and by the School of Physics at the University of Melbourne.

\newpage
\begin{figure}
\caption{The temparature of the critically fragmenting disc as a function
of the orbital period.}
\end{figure}
\begin{figure}
\caption{The surface density of the critically fragmenting disc as
a function of the orbital period.}
\end{figure}
\begin{figure}
\caption{The Toomre mass of the critically fragmenting disc as
a function of the orbital period.}
\end{figure}
\begin{figure}
\caption{The isolation and gap opening masses plotted
as a function of radius for the critically fragmenting disc without
external sources of heating. The black hole mass is taken to be
$3\times 10^6M_{\odot}$.}
\end{figure}

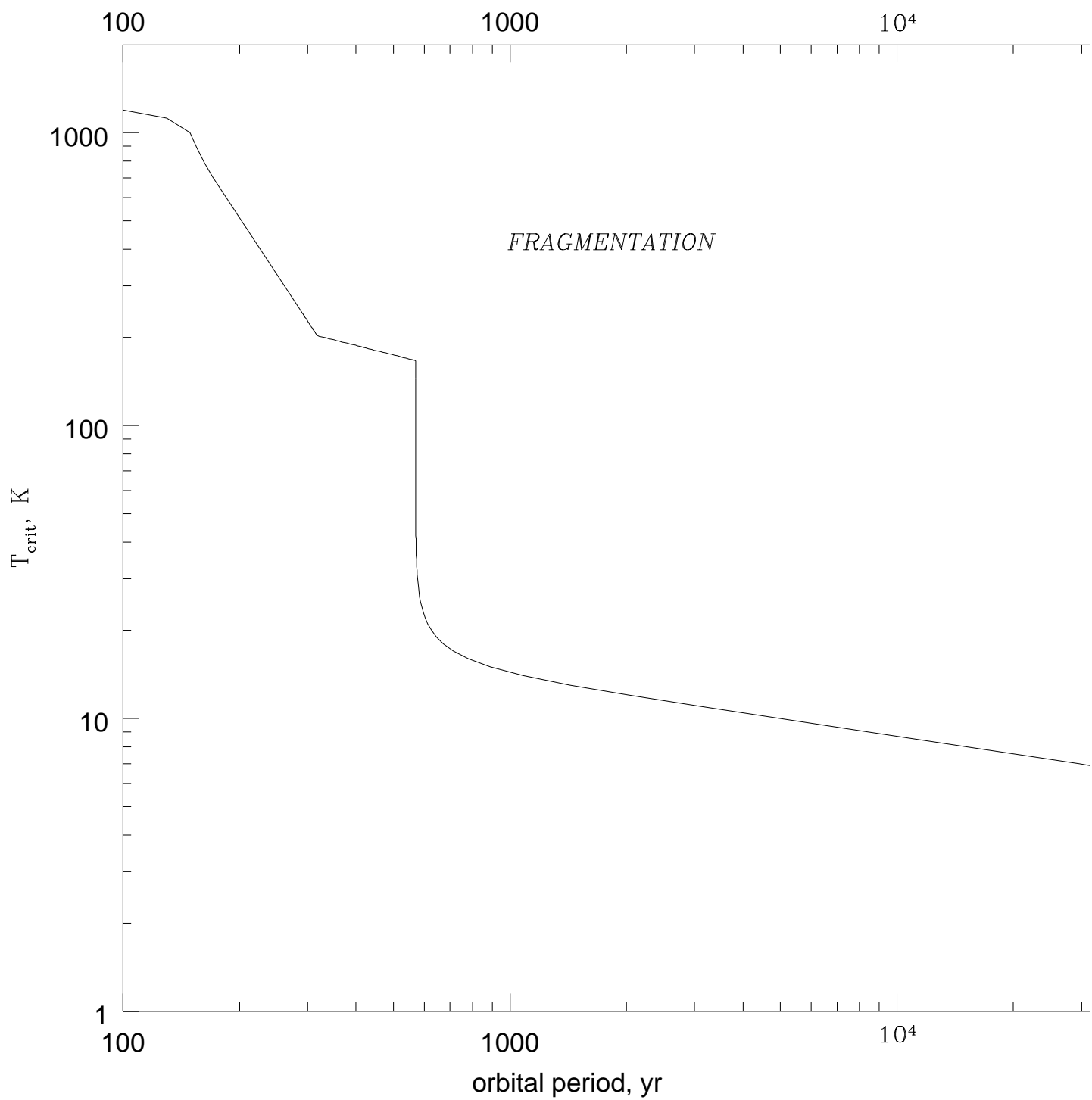
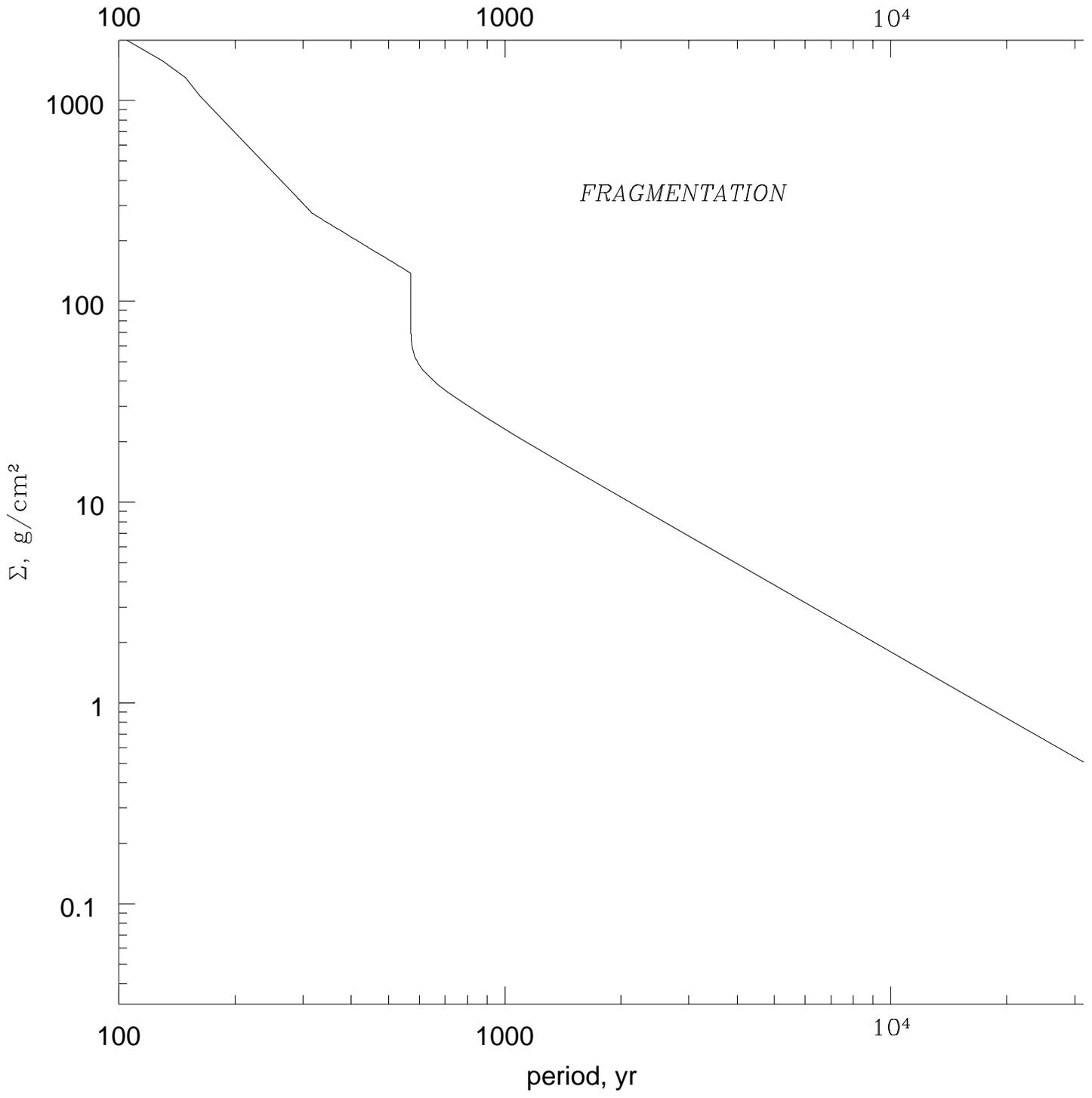
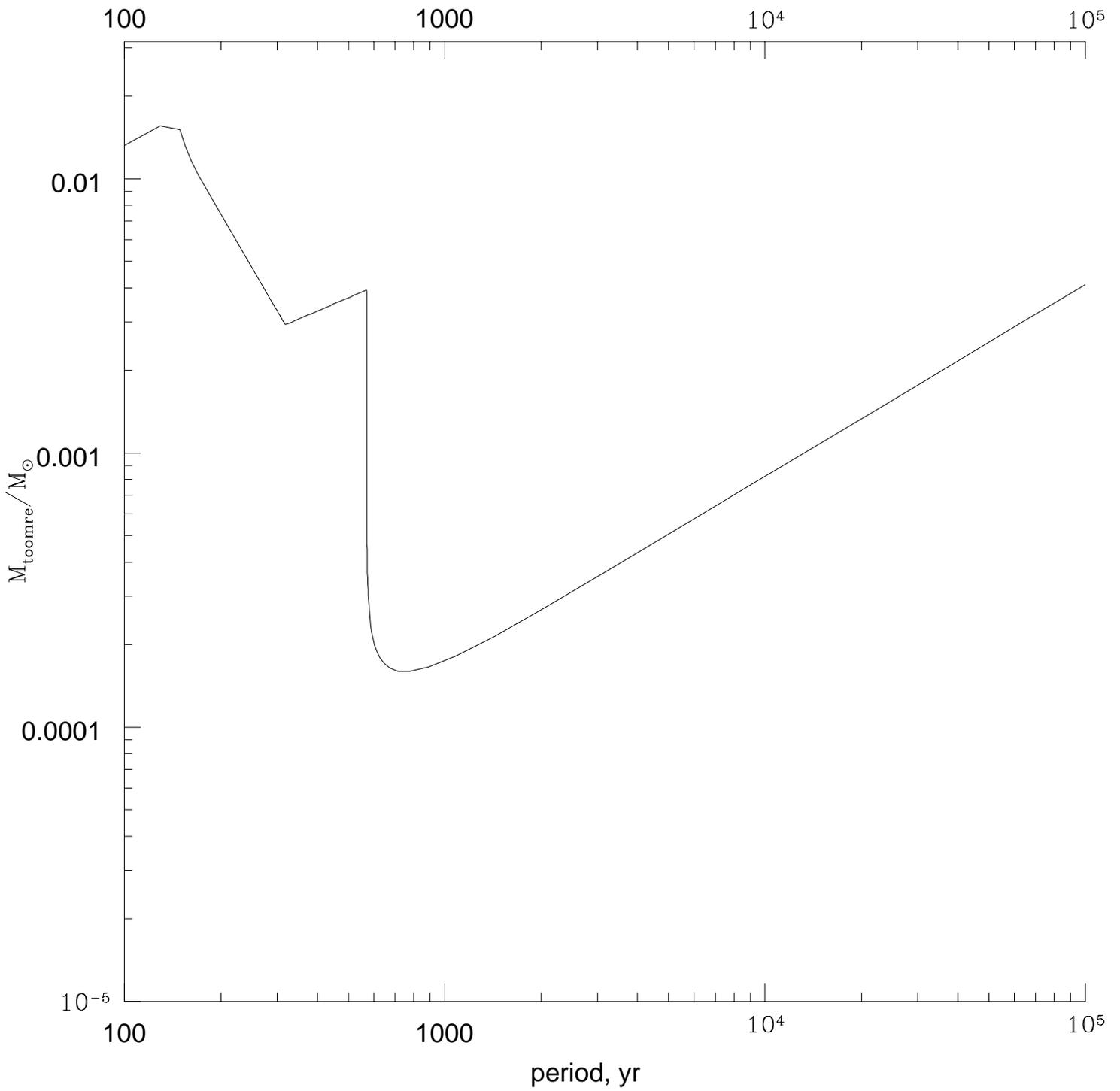
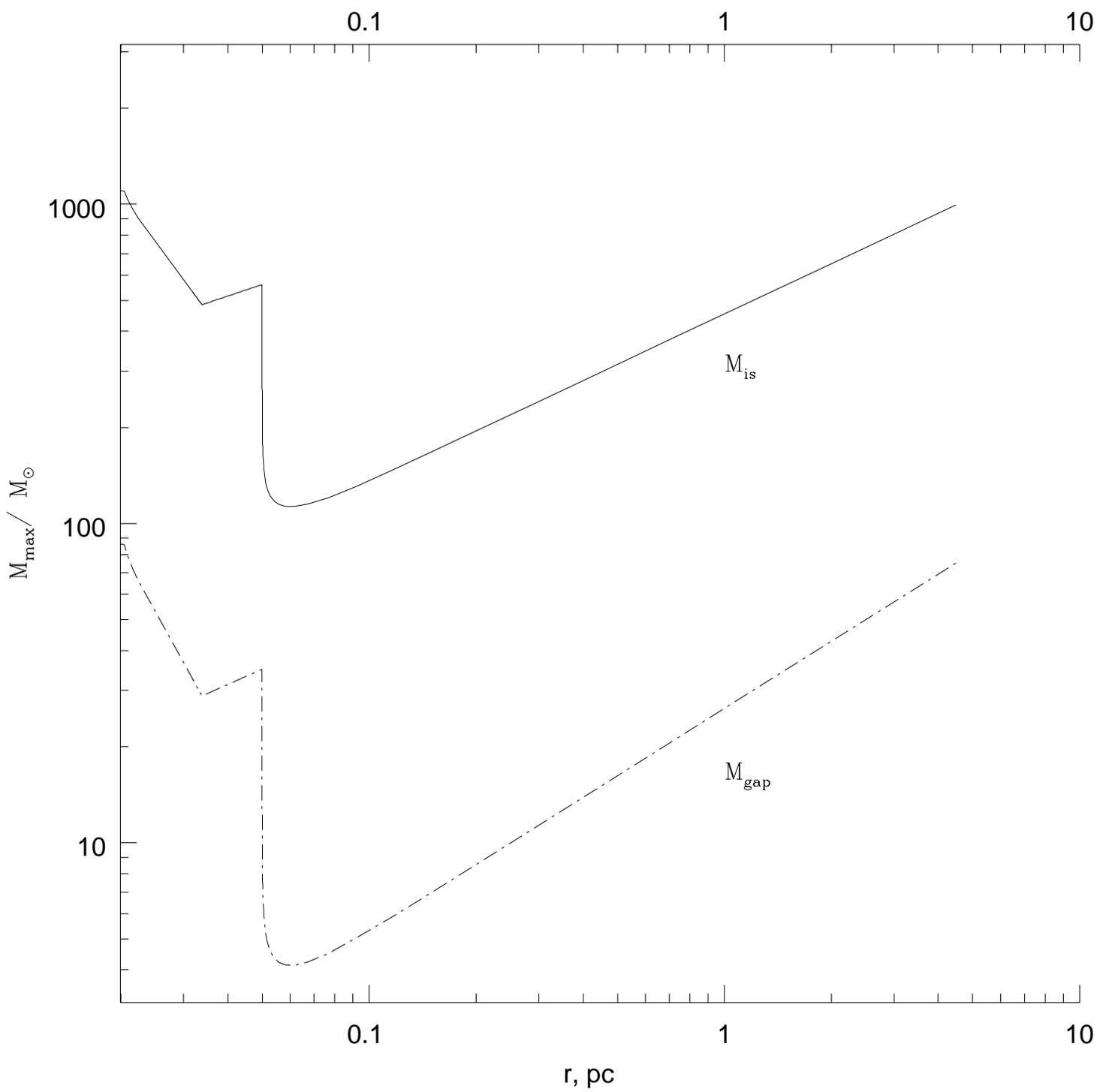
\end{document}